# Analytical and Cross-Sectional Clinical Validity of a Smartphone-Based U-Turn Test in Multiple Sclerosis

Marta Płonka,[a*] Rafał Klimas,[b*] Dimitar Stanev,[b] Lorenza Angelini,[b] Natan Napiórkowski,[a] Gabriela González Chan,[c] Lisa Bunn,[c] Paul S Glazier,[c] Richard Hosking,[d] Jenny Freeman,[c] Jeremy Hobart,[e] Mattia Zanon,[b**] Jonathan Marsden,[c**] Licinio Craveiro,[b**] and Mike D Rinderknecht[b**]

*shared first authorship

**shared last authorship

[a]Roche Polska Sp. z o.o., Warsaw, Poland

[b]F. Hoffmann-La Roche Ltd, Basel, Switzerland

[c]School of Health Professions, University of Plymouth, Plymouth, UK

[d]University of Plymouth Enterprise Ltd, Plymouth, UK

[e]Peninsula Medical School, Faculty of Health, University of Plymouth, Plymouth, UK

**Corresponding author:**

Mike D Rinderknecht
mike.rinderknecht@roche.com

# Abstract

**Background:** Gait and balance impairment can profoundly impact people with multiple sclerosis (PwMS).

**Objectives:** To evaluate the analytical and clinical validity of the U-Turn Test (UTT), a smartphone-based assessment of dynamic balance in PwMS.

**Methods:** The GaitLab study (ISRCTN15993728) enrolled adult PwMS (EDSS 0.0–6.5). PwMS performed the UTT in a gait laboratory (supervised) using 6 smartphones at different wear locations and daily during a two-week remote period (unsupervised) using one smartphone (belt front). Median turn speed was computed per UTT. In the supervised setting, turn detection accuracy of smartphones was compared to motion capture (mocap) via F1 scores. Agreement between smartphone- and mocap-derived turn speed was assessed by Bland-Altman and ICC(3,1). In the unsupervised setting, test-retest reliability (ICC[2,1]) and correlations with Timed 25-Foot Walk (T25FW), EDSS, Ambulation Score, 12-item Multiple Sclerosis Walking Scale (MSWS-12), and Activities-specific Balance Confidence scale (ABC) were evaluated.

**Results:** Ninety-six PwMS were included. Turn speed was comparable across supervised (1.44 rad/s) and unsupervised settings (1.47 rad/s). In the supervised setting, turn detection was highly accurate (F1 >95% across wear locations). Turn speed agreement with mocap was high (ICC[3,1]: 0.87–0.92), with minimal bias (-0.04 to 0.11 rad/s). Unsupervised test-retest reliability (ICC[2,1]) was >0.90 when aggregating ≥2 tests. Turn speed correlated with T25FW (rho=-0.79), EDSS (rho=-0.75), Ambulation score (rho=-0.73), MSWS-12 (rho=-0.65), and ABC (rho=-0.61).

**Conclusion:** The UTT accurately and reproducibly measures turn speed across wear locations and settings, providing complementary dynamic balance insights to clinical measures and showing potential for use in multiple sclerosis trials.

**Glossary:**

ABC, Activities-specific Balance Confidence scale

EDSS, Expanded Disability Status Scale

ICC, intraclass correlation coefficient

IQR, interquartile range

MS, multiple sclerosis

MSWS-12, 12-item Multiple Sclerosis Walking Scale

PwMS, people with multiple sclerosis

T25FW, Timed 25-Foot Walk

UTT, U-Turn Test

# 1. Introduction

Multiple sclerosis (MS) affects approximately 2.8 million people worldwide and is the leading cause of non-traumatic disability among young adults (Walton et al. 2020). It can cause impairment across multiple functional domains, including gait and balance (Thompson et al. 2018a), with approximately two-thirds of people with MS (PwMS) reporting problems with balance and coordination (Prosperini and Castelli 2018).

Gait and balance impairment have a profound impact on affected individuals (LaRocca 2011; Yildiz 2012), including a significantly increased risk of falls, with patients reporting at least one fall over a 3-6–month period (Cattaneo et al 2002; Coote et al. 2013; Finlayson et al. 2006). Falls that occur during turning are particularly hazardous. A study in a geriatric, non-MS population reported that falls while turning are associated with an almost 8 times greater risk of hip fractures than falls while walking in a straight line (Cumming and Klineberg 1994). Given that turning is a crucial component of daily activities and accounts for about one-third of daily steps involving changes in walking direction (Glaister et al. 2007), the ability to effectively adapt gait and balance for safe turns is vital for maintaining safety and independence (Robinson and Ng 2018). Examining turns can also provide insights into early impairments of dynamic balance—which involves controlling balance and posture while varying the base of support as well as moving the center of mass over the base of support (Petró et al. 2017; Rizzato et al. 2025)—, as turning is more complex and demanding than straight walking. Turning requires greater neural resources for planning and coordinating postural transitions than walking in a straight line, increased integration between balance and gait control systems, and enhanced spatial coordination among limbs (Herman et al. 2011; Kling et al. 2012). Unsurprisingly, research suggests that turning may be more vulnerable to impairment than straight walking (Crenna et al. 2007; Zampieri et al. 2010).

Different tools are available to assess dynamic balance, including the Timed Up and Go Test (Sebastião et al. 2016) and the Berg Balance Scale (Cattaneo et al. 2006). These assessments are typically administered by clinicians in a healthcare setting, limiting their use to clinic visits that occur infrequently and in an environment that does not reflect everyday functioning. While testing in supervised settings ensures a standardized and controlled environment for a test, remote, unsupervised assessments that are performed at home without supervision by clinical staff enable a much higher testing frequency resulting in a more detailed picture of the patient's daily functioning.

By leveraging sensors embedded in commercially available smartphones, the U-Turn Test (UTT) has been proposed as a simple and convenient tool for remotely assessing dynamic balance at home without any supervision (Cheng et al. 2011; Montalban et al. 2022). The UTT involves walking back and forth for 60 s between two points a few meters apart, making a U-turn each time one of the points is reached. In a proof-of-concept study, turn speed derived from the UTT correlated with clinical assessments such as EDSS ($\rho$ = -0.45) or performance tests such as T25FW ($\rho$ = -0.51), while showing excellent test-retest reliability using UTT measures aggregated across two-week periods (intraclass correlation coefficient [ICC[2,1]] = 0.83 – 0.87) (Cheng et al. 2011; Montalban et al. 2022). The GaitLab study (ISRCTN15993728) (Rinderknecht et al. 2023) was designed to develop further new or existing algorithms that detect turns and derive improved digital measures that characterize turning patterns. Here we aim to validate the technical and clinical performance of these improved algorithms by comparing the UTT against a gold-standard motion capture system and by determining correlations with established clinical scales in a population of patients with MS. The following questions are addressed: (I) How accurately does the turn detection algorithm detect U-turns in smartphone sensor data; (II) How do smartphone-derived UTT measures relate to reference turn measures, and how are they impacted by different smartphone wear locations; (III) how reproducible are smartphone-derived UTT measures in unsupervised, real-world setting; and (IV) how do smartphone-derived UTT data relate to other clinical indicators of gait or walking in a broad MS population.

# 2. Methods

## 2.1. Study design and participants

The GaitLab study was designed to evaluate the measurement properties of smartphone-based digital measures of gait and balance. The study design has been previously described in detail (Rinderknecht et al. 2023) and is summarized in Figure 1. In short, this single-site study included two on-site visits in Plymouth, UK and a two-week remote, unsupervised testing period in between. Full neurological and clinical assessments were performed during the first on-site visit. These included the EDSS (Kappos 2011), Ambulation Score (a subscore of the EDSS) (Kappos 2011), Timed 25-Foot Walk (T25FW) (Motl et al. 2017) and the Falls Questionnaire (items include "Have you fallen in the last 3 months?" and "If yes, how many times during this period did you fall? Once? Twice? Three times or more?"). During the remote testing period, all study participants performed daily, unsupervised, smartphone-based UTT in their home environment. Additionally, they were tasked to complete once during this period patient-reported outcomes, including the 12-item Multiple Sclerosis Walking Scale (MSWS-12) (Hobart et al. 2003) and the Activities-specific Balance Confidence scale (ABC) (Myers et al. 1998). Finally, during the second on-site visit, all participants performed the UTT in a motion-capture laboratory.

The study enrolled both PwMS and healthy controls (HC) from a single site in Plymouth, United Kingdom, between January 2022 and October 2023. All PwMS were ≥18 years old, and had a confirmed diagnosis of either relapsing-remitting, primary progressive, or secondary progressive MS according to the 2010 or 2017 McDonald criteria (Thompson et al. 2018b), and a baseline EDSS between 0.0 (no neurological deficits) and 6.5 (bilateral assisted walking).

The study was approved by the UK Health Research Authority (IRAS Project ID: 302099), and all study participants provided written informed consent prior to any study procedures.

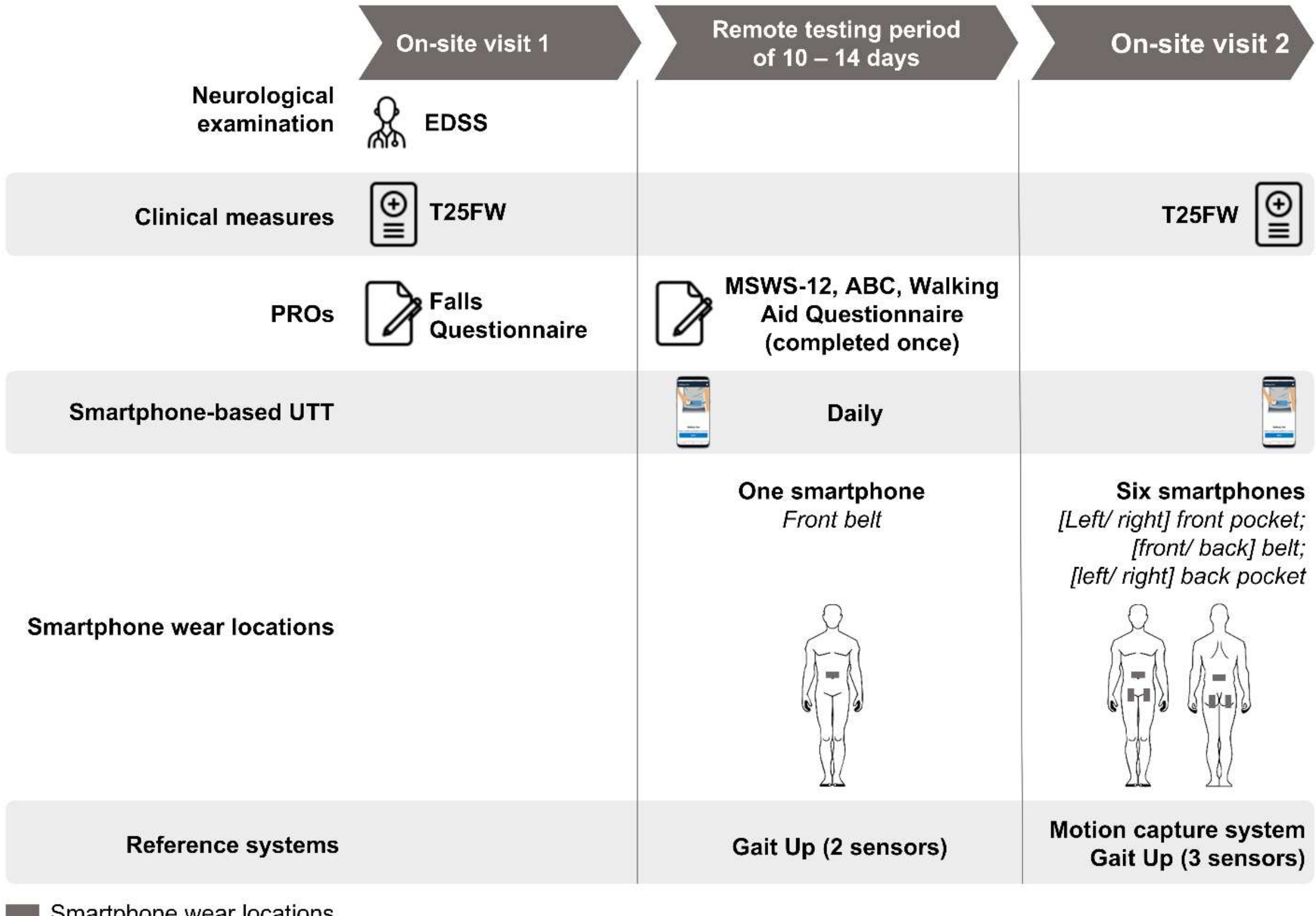

**Figure 1. Design of the GaitLab study.** UTT data collected during the remote testing period and the second on-site visit were used in the analyses.

ABC, Activities-specific Balance Confidence Scale; EDSS, Expanded Disability Status Scale; MSWS-12, 12-item Multiple Sclerosis Walking Scale; PRO, patient-reported outcomes; T25FW, Timed 25-Foot Walk; UTT, U-Turn Test.

## 2.2. U-Turn Test

The UTT is a smartphone-based test aimed to assess gait and dynamic balance in PwMS in both supervised and unsupervised settings (Cheng et al. 2011; Montalban et al. 2022). The UTT was performed during the two on-site visits in a gait laboratory with motion capture capabilities under the supervision of clinical staff (data from only the second on-site visit were used). During these sessions, participants were instructed to walk safely at their normal walking speed for 5 meters before making a U-Turn, and to continue to do so for 60 seconds. Inertial measurement unit (IMU) data were simultaneously recorded with six smartphones placed in six body locations: each left and right front and back trouser pockets, and back and front of a waist-worn belt (Figure 1). While sensors strapped to the waist are more commonly used in gait studies (Karatsidis et al. 2025), carrying the smartphone in a garment (e.g., trouser) pocket reflects more the real-world environment (Redmayne 2017). Thus, incorporating these different smartphone wear locations enabled us to study how digital measures derived from the UTT are influenced by smartphone placement and whether different smartphone placements could be supported. We re-labeled these locations based on the role of the leg during the turn, whether the leg was on the outer or inner side relative to the turn direction (inner front/back pockets, back belt, front belt, and outer front/back pockets). For example, during left turns, the smartphone location labels for pockets on the right leg are re-labeled as outer front or back pocket.

In the unsupervised setting, the UTT was performed remotely in the study participants' home environment once daily during the two-week remote testing period using a single smartphone device worn in a belt bag at the waist level (front belt) to reduce the burden on the study participants. To perform this test, the study participants were instructed to walk back and forth for 60 s between two points a few meters apart safely and at their normal speed, making a U-turn each time they reached one of the points.

The participants were allowed to use walking aids and/or orthotics and rest as needed in both the supervised and unsupervised settings, both for safety reasons and for capturing their real-world gait. Additionally, the first supervised on-site visit served as a functional safety screening, during which a member of the site staff determined the feasibility and safety of subsequent unsupervised remote sessions for each participant. During the second on-site visit, staff proactively mitigate fall risk if necessary (Rinderknecht et al. 2023).

### 2.2.1. Digital measures derived from the U-Turn Test

Smartphone-derived turn speed was computed as the estimated turn angle divided by turn duration (descriptive statistics of number of turns and turn duration are provided in Supplementary Tables S1 and S2). Deriving these measures involved two steps: 1) automatic detection of turn bouts (a period of uninterrupted turning while walking) with a turn detection algorithm and 2) computation of digital measures obtained during these bouts. The turn detection algorithm is based on the algorithm of El-Gohary et al. (El-Ghohary et al. 2013). As El-Ghohary et al.'s algorithm was designed to detect turns of 45 deg rather than the 180 deg turns of our UTT, some modifications to their algorithm were necessary. For example, the threshold for the parameter minimum rotational rate on the vertical axis was increased from 5 to 20 deg/s. Preliminary investigations showed that a lower threshold of 5 deg/s triggered spurious turns due to pelvis rotations and leg swing artifacts during walking. Therefore, a higher threshold was selected to make the detection more robust. The 20 deg/s was chosen by inspecting the range of rotation during straight walking such that rotations are less likely to trigger turns accidentally. Additionally, the threshold for a minimal turn angle was increased to 90 deg.

Using this modified turn detection algorithm, individual U-turn bouts were automatically detected in the smartphone sensor data. The sensor data were recorded using a smartphone application (formerly referred to as Floodlight) that was installed on Samsung Galaxy A40 smartphones running Android version 9. Sensor data from the in-built IMU sensors (triaxial accelerometer, gyroscope, and magnetometer sensors) were collected at a sampling frequency of 50 Hz, allowing detection of most human body motion frequencies (Antonsson and Mann 1985; Zeng and Zhao 2011). The accelerometer had a range of ± 4 g and a resolution of 0.122 mg/LSB, while the gyroscope had a range of ± 1000 deg/s and a resolution of 30.5 mdeg/s/LSB.

## 2.3. Reference measurement systems

Temporal synchronization of data from reference measurement systems and smartphones was ensured via cross-correlation before the laboratory protocol started (Stanev et al. 2024). The reference measurement system used in the supervised setting was a motion-capture system. It included 12 infrared cameras (Vicon Vero™ v1.3) and 26 reflective markers (14 mm in diameter) positioned on body landmarks based on the Plug-in Gait marker set (note that additional markers were placed on the smartphones and some markers on the upper body were excluded). Marker trajectories were recorded at a sampling frequency of 100 Hz, and key gait cycle events such as initial contacts (i.e., heel strike) and final contacts (i.e., toe off) were extracted with the Gait Offline Analysis Tool software (Motek, Netherlands). Following the procedure of Rehman et al. (Rehman et al. 2020), the beginning and end of the turns in the motion-capture trajectories ("reference turns") were identified by four raters with extensive biomechanics expertise (JM, GGC, LB, PSG) based on observations of the gait marker movement simulation (stick figure) and added as events at specific Vicon frames. To ensure consistency in identifying turns, motion capture data were flagged for further inspection if the number of turns did not match the number of turns detected with the second reference measurement system (i.e., Gait Up IMUs). The beginning of the turn was defined as the time of initial contact of the first step in the turn, while the end of the turn was defined by the time of final contact of the last step. After identification of turns, turn measures were derived for each turn and subsequently aggregated as the median of all turns detected during a single UTT. As data on turn angle were not available for the motion capture system, the turn angle was assumed to be π radians (= 180 degrees), and turn speed was computed as π / turn duration (rad/s). The validity of this assumption was confirmed by the smartphone-derived measurements, which yielded a median turn angle ranging from 3.0 – 3.2 rad, demonstrating that participants adhered closely to the 180-degree instruction.

Gait Up IMUs (Physilog 6® model, Gait Up, Lausanne, Switzerland) were the second reference measurement system deployed in the study, which consist of body-worn wearable units with triaxial accelerometer (range: ± 8 g; sensitivity: 0.244 mg/LSB), gyroscope (range: ± 2000 deg/s; sensitivity: 70 mdeg/s/LSB), and magnetometer (range: ± 50 mT; resolution: 0.161 – 3.22 μT/LSB [along the x- and y-axes] and 0.294 – 5.87 μT/LSB [along the z-axis]) sensors that acquire data with a sampling frequency of 128 Hz. A three-sensor set-up was used supervised setting, with one sensor placed on each foot and one additional sensor on the lower back. In the unsupervised setting, a simpler set-up was used, with two sensors placed on each foot for ease of use and minimizing participant burden.

## 2.4. Statistical analysis

### 2.4.1. Validation of turn detection algorithm

To validate the smartphone-based turn detection algorithm, it was assessed whether turns detected by the smartphone coincided temporally with turns detected by the reference measurement system (i.e., the motion capture system). But given that the turn detection algorithm and the reference measurement capture system use different criteria to define a turn, it was considered sufficient if a turn detected by the smartphone overlapped with at least 20% of the corresponding reference turn (green shading in Figure 2). A minimum of 20% overlap was chosen to ensure that the overlap exceeds the minimum turn duration accepted by the smartphone's turn detection algorithm (minimally accepted turn duration of 0.5 s = 20% of the mean reference turn duration of 2.6 s). This 20% threshold, derived from interim data analyses, allows matching turns detected with the smartphone to corresponding reference turns without excluding valid turns due to sub-second timing shifts. Such shifts may result from inherent differences in turn definitions used by the reference measurement system and the smartphone-based turn detection algorithm, and synchronization of the two measurement systems.

Individual turns detected by the smartphone were classified as either correctly detected (true positive [TP]; smartphone and reference turn overlapped temporally by ≥ 20%), incorrectly detected (false positive [FP]; smartphone, but not reference measurement system, detected a turn, or the temporal overlap is < 20%), or not detected (false negative [FN]; reference measurement system, but not smartphone, detected a turn) (Figure 2). The performance of the turn detection algorithm was evaluated across all turns of a UTT. Performance measures included recall (i.e., sensitivity), precision (i.e., positive predictive value) and F1 score (harmonic mean of precision and recall):

$$Recall = \frac{TP}{TP + FN}$$

$$Precision = \frac{TP}{TP + FP}$$

$$F1\ score = \frac{2 * (precision * recall)}{precision + recall}$$

Precision represents the proportion of true positives among all events detected by the algorithm. Recall, or sensitivity, by comparison, is the proportion of true positives among all events that should have been detected. Finally, the F1 score is the harmonic mean of precision and recall and provides a balanced measure of the turn detection algorithm's performance (Taha and Hanbury 2015).

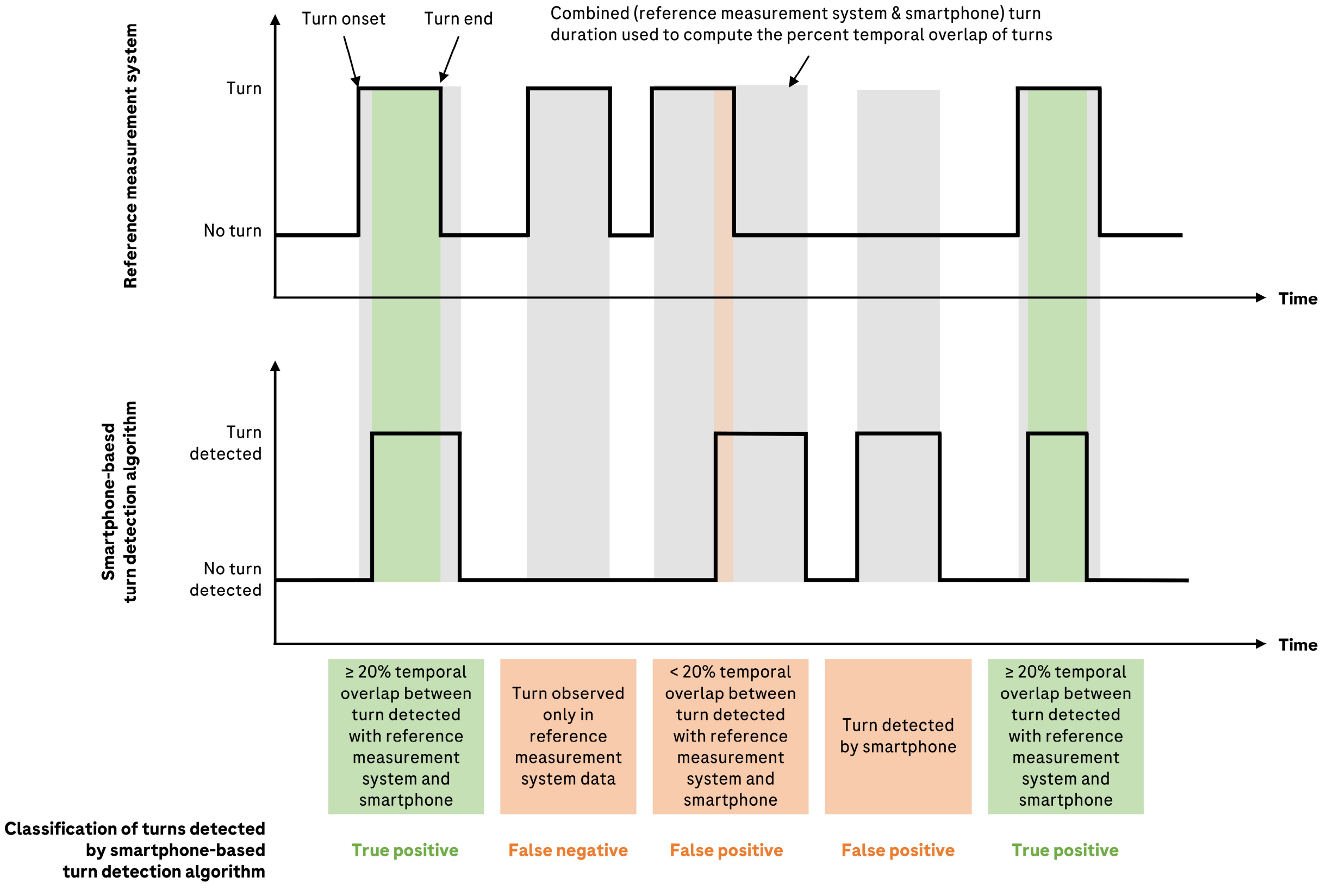

**Figure 2. Classification of turns for validation of turn detection algorithm.** Classification of turns detected by the smartphone-based turn detection algorithm is based on the temporal overlap between turns observed in the reference measurement system data and turns detected by the smartphone-based turn detection algorithm. Turns observed, or detected, by both measurement systems were classified as either true positive (≥ 20% temporal overlap) or false positive (< 20% temporal overlap) depending on how much the respective turn overlapped temporally. Turns observed, or detected, by only one measurement system were classified either as false negative (observed in reference measurement system data only) or false positive (detected by the smartphone-based turn detection algorithm only).

In addition, the temporal errors, or temporal differences, between the onset (or end) of turns detected by the turn detection algorithm and by the reference measurement system were computed. These temporal errors were evaluated for all correctly classified turns, i.e., all true positives. A negative temporal error indicates the smartphone-based turn had started (or ended) before the turn was detected by the reference system. Conversely, a positive temporal error indicates that the smartphone-based turn had started (or ended) after the reference turn.

Both analyses were performed for each of the six smartphone wear locations using data collected in the supervised setting.

### 2.4.2. Validation of turn speed median

Measures from individual turns performed during each UTT were aggregated by taking the median, resulting in the measure turn speed median. The validation consisted of the following steps:

*Step 1 – Agreement between smartphone and reference measurement system*

Agreement was assessed for each of the six smartphone locations included in the supervised setting by comparing the turn speed median against turn speed median measured with the reference system. Absolute agreement with 95% confidence intervals was evaluated with intraclass correlation coefficients (ICC[3,1]) [36]. The ICC(3,1) was calculated based on a 2-way mixed effects model, treating the smartphone and reference measurement systems as two specific raters. This model was selected because these systems are not assumed to be representative of, or generalizable to, all other possible measurement algorithms or motion-tracking systems. ICCs(3,1) < 0.50 were considered as poor, ICCs(3,1) between 0.50 and < 0.75 as moderate, ICCs(3,1) between 0.75 and < 0.90 as good, and ICCs(3,1) ≥ 0.90 as excellent (Koo and Li 2016). Bland-Altman plots and respective metrics, including bias between the mean differences and the limits of agreements (LoA) were computed (Bland and Altman 1999). Confidence intervals (95% CI) for bias and LoA were obtained through a bootstrapping approach with randomized sampling of participants and 500 repetitions.

Due to the suboptimal agreement between the silver-standard Gait Up IMU sensors and the gold-standard motion capture system (see supplementary appendix p. 2), the Gait Up sensors could not reliably be used as a reference system. For this reason, this analysis was conducted on data collected in the supervised setting with the motion capture system as the reference measurement system.

*Step 2 – Consistency of smartphone-based algorithms across repeated assessments*

In scenarios where high-frequency testing is possible, such as in the remote testing period, measurement consistency may be improved, and measurement error may be reduced by aggregating multiple tests taken over a relatively short period of time. The reliability or reproducibility was, therefore, assessed using ICC(2,1) for absolute agreement as a function of the number of UTTs performed in the unsupervised setting (note: the individual UTTs are regarded as a random rater representative of multiple raters as required by the ICC[2,1]) (Koo and Li 2016). This allows establishing the minimum number of UTTs that need to be aggregated to achieve an acceptable level of reproducibility and measurement error. An ICC(2,1) < 0.50 was considered as being poor, an ICC(2,1) between 0.50 and < 0.75 as moderate, an ICC(2,1) between 0.75 and 0.90 as good, and an ICC(2,1) > 0.90 as excellent [36]. Additional metrics include the standard error of measurement (SEM) and the minimally detectable change (MDC), including their 95% CI. The SEM is defined as $SEM = \sqrt{var_{\varepsilon}}$ , where $var_{\varepsilon}$ is the within-patient residual variance, and the MDC as $MDC = 1.96 * SEM * \sqrt{2}$.

All metrics were estimated using mixed-effect model variances. Confidence intervals were obtained using bootstrapping with randomized sampling of participants and randomized selection of the number of test repetitions. This bootstrapping approach was conducted with 500 repetitions.

### *Step 3 – Correlation of smartphone-based measures with clinical and patient-reported measures*

Clinical validity was evaluated by correlating the smartphone-derived turn speed obtained during the unsupervised, remote period with clinical and patient-reported measures. In a first step, measurements of turn speed median obtained from individual tests were aggregated on a participant level by taking the median of daily values across the two-week period. This single aggregated measure of turn speed median was correlated with the following clinical scales and patient-reported outcomes using Spearman rank correlation: EDSS, Ambulation Score, MSWS-12, T25FW-derived gait speed, and ABC. The clinical scales were administered once during the on-site visits, and patient-reported outcomes were completed by participants during the remote testing period. The correlation strength was considered as follows: $|\rho|$: 0.80 – 1.00 = very strong; $|\rho|$: 0.60 – 0.79 = strong; $|\rho|$: 0.40 – 0.59 = moderate; $|\rho|$: 0.20 – 0.39 = weak; $|\rho|$: 0.00 – 0.19 = very weak (Swinscow 1997).

The ability of turn speed median measured during the unsupervised, remote testing period to differentiate between groups of PwMS was also determined. The groups were defined based on the EDSS (0–3.5 vs 4–5.5 vs 6–6.5), Ambulation score (0 – 1 vs 2 – 5 vs 6 – 9), Falls Questionnaire (≥ 1 fall in the 3 months prior to baseline vs no fall in the 3 months prior to baseline), and use of walking aids (yes vs no). Data was represented with boxplots and differences between groups were assessed for statistical significance using the Mann-Whitney test U. P-values $< 0.05$ were considered as statistically significant.

# 3. Results

## 3.1. Population and demographics

The GaitLab study enrolled 100 patients with MS. Data from 96 PwMS were available for validating the turn detection algorithm, 93 PwMS for validating turn speed median in the supervised setting, and 91 PwMS for validating turn speed median in the unsupervised setting (see Figure 3 for a detailed patient disposition). Baseline demographics and disease characteristics of the 96 PwMS included in the validation analysis of the turn detection algorithm are presented in Table 1.

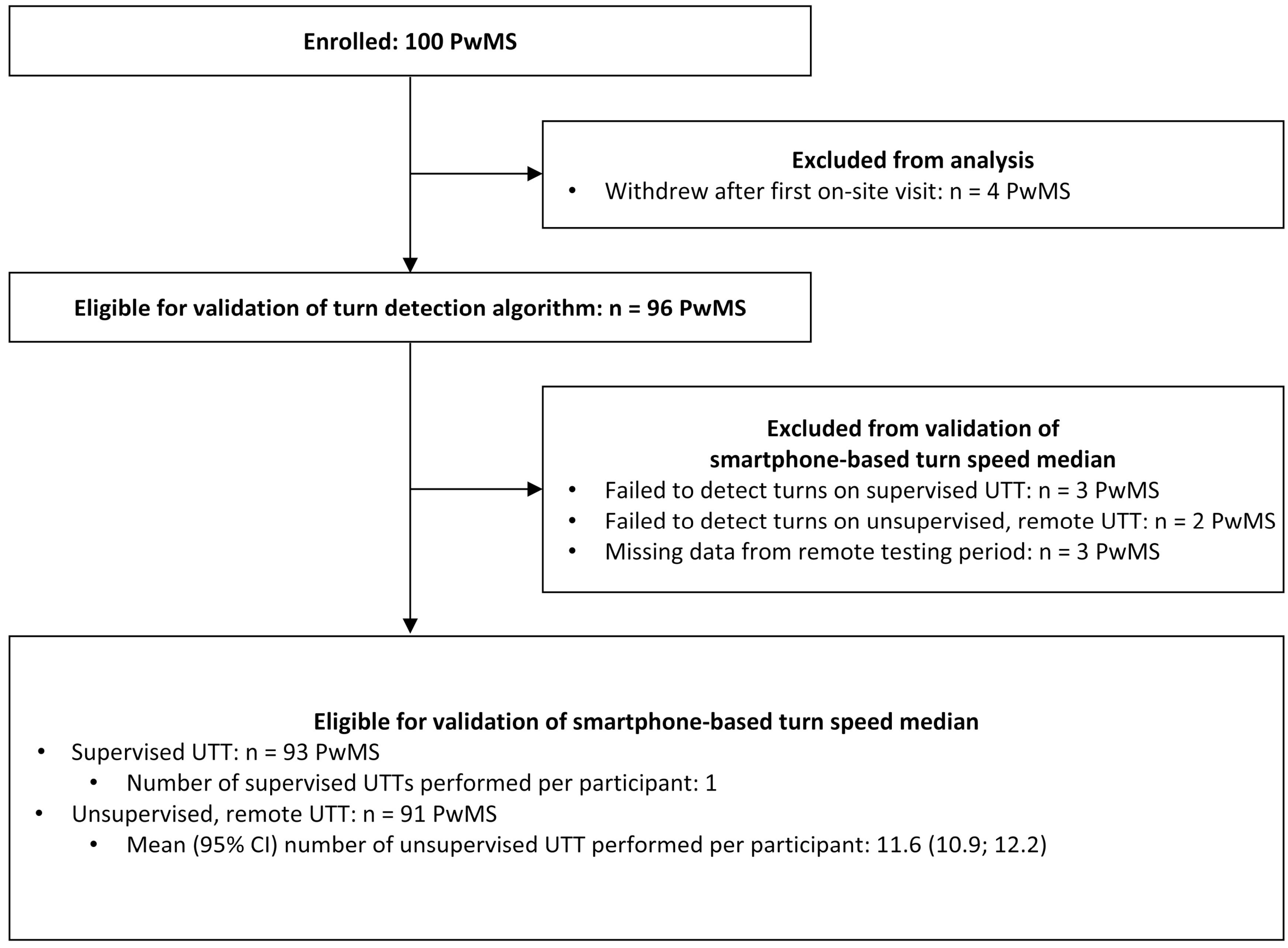

**Figure 3. Participant flowchart.**

PwMS, people with multiple sclerosis; UTT, U-Turn Test.

**Table 1. Baseline demographics and disease characteristics.**

| | PwMS (n=96) |
|---|---|
| Age (years), median (IQR); range | 56.0 (46.8 – 63.2); 26 – 77 |
| Female, n (%) | 59 (61.5) |
| Height (m), median (IQR); range | 1.7 (1.6 – 1.8); 1.5 – 1.9 |
| Weight (kg), median (IQR); range | 75.0 (64.8 – 93.0); 46.0 – 142 |
| MS type, n (%) | |
| Relapsing-remitting | 67 (69.8) |
| Secondary progressive | 16 (16.7) |
| Primary progressive | 13 (13.5) |
| Time since MS symptom onset (years), median (IQR); range | 18 (11 – 27); 3 – 53 |
| T25FW (s), median (IQR); range | 5.4 (4.5 – 8.4); 2.9 – 40.2 |
| Ambulation score, median (IQR); range | 3.5 (1.0 – 5.0); 0.0 – 9.0 |
| MSWS-12 v2 transformed total score, median (IQR); range | 47.0 (21.0 – 73.0); 0.0 – 100.0 |
| EDSS | |
| Median (IQR); range | 5.3 (4.0 – 6.0); 1.5 – 6.5 |
| 0 to 3.5, n (%) | 23 (24.0) |
| 4.0 to 5.5, n (%) | 35 (36.5) |
| 6.0 to 6.5, n (%) | 38 (39.6) |
| ABC, median (IQR); range | 68.1 (44.1 – 85.6); 6.2 – 100.0 |
| PwMS with falls in the last 3 months, n (%) | 34 (35.4) |
| Use of walking aids, n (%) | 60 (62) |

ABC, Activities-specific Balance Confidence scale; EDSS, Expanded Disability Status Scale; IQR, interquartile range; MS, multiple sclerosis; MSWS-12, 12-item Multiple Sclerosis Walking Scale; PwMS, people with multiple sclerosis; T25FW, Timed 25-Foot Walk.

## 3.2. Validation of turn detection algorithm

In the supervised setting, the smartphone-based algorithm detected a median (range) of 12 (0 – 22) turns compared to 11 (1 – 21) turns detected with the reference motion capture system. The number of detected turns was similar across the six different smartphone locations (for more details see Supplementary Table S1). Most turns were correctly detected (true positives), and a very low number of turns were incorrectly detected (false positives), translating into excellent precision (mean values ≥ 96.1%; Supplementary Table S4). The number of turns identified by the reference system which were not detected by the smartphone system (false negatives) was also very low, translating into an excellent recall or sensitivity (mean values ≥ 95.2%; Supplementary Table S5). The combination of the two metrics was also excellent with F1 scores exceeding 95%. Importantly, no differences were observed across the six tested smartphone locations, with each location achieving a mean F1 score of > 95% (Table 2). The lowest F1, recall and precision scores were observed in groups with greater disability (PwMS with EDSS 6.0 – 6.5), but the turn detection algorithm still performed on a high level in most participants in this subgroup (Supplementary Tables S6 – S8).

The correct detection of turns required an overlap of at least 20% of the turn time between smartphone and reference system. Mean temporal errors between smartphone-based turns and reference turns were minimal, ranging from -0.01 s to 0.05 s for the start time, and from -0.06 s to -0.02 s for the end time (Table 3). The smartphone algorithm, therefore, detected turns with high temporal precision, with a trend towards a minor earlier end time relative to the reference system. The mean overall temporal overlap was 80.4%. Results were consistent across all six smartphone wear locations (Table 3).

**Table 2. F1 score statistics for turn detection algorithm applied during the supervised setting.**

| Smartphone wear location | Distribution of F1 scores across PwMS[a] | | | | | | | | |
|---|---|---|---|---|---|---|---|---|---|
| | Mean (95% CI) | SD | Min | Max | P05 | Q1 | Median | Q3 | P95 |
| Pocket front outer | 95.5 (91.9; 99.2) | 18.3 | 0.0 | 100 | 80.0 | 100 | 100 | 100 | 100 |
| Pocket back outer | 95.8 (92.2; 99.3) | 17.9 | 0.0 | 100 | 84.3 | 100 | 100 | 100 | 100 |
| Belt front | 95.7 (92.1; 99.3) | 17.9 | 0.0 | 100 | 84.3 | 100 | 100 | 100 | 100 |
| Belt back | 96.0 (92.4; 99.5) | 17.8 | 0.0 | 100 | 88.7 | 100 | 100 | 100 | 100 |
| Pocket front inner | 95.3 (91.7; 99.0) | 18.5 | 0.0 | 100 | 76.7 | 100 | 100 | 100 | 100 |
| Pocket back inner | 95.7 (92.1; 99.3) | 17.9 | 0.0 | 100 | 84.3 | 100 | 100 | 100 | 100 |

[a]F1 scores are reported as a percentage. CI, confidence intervals; SD, standard deviation; min, minimum value; max, maximum value; P05, 5th percentile; Q1, 1st quartile; Q3, 3rd quartile; P95, 95th percentile, PwMS, people with multiple sclerosis.

**Table 3. Temporal errors in the supervised setting between turns detected by the turn detection algorithm and reference measurement system.[a]**

| Smartphone wear location | Temporal error for turn onset [s] | | Temporal error for turn end [s] | |
|---|---|---|---|---|
| | Mean | SD | Mean | SD |
| Pocket front outer | 0.05 | 0.4 | -0.06 | 0.4 |
| Pocket back outer | -0.01 | 0.3 | -0.04 | 0.4 |
| Belt front | 0.00 | 0.3 | -0.05 | 0.5 |
| Belt back | 0.00 | 0.3 | -0.05 | 0.5 |
| Pocket front inner | -0.01 | 0.3 | -0.02 | 0.5 |
| Pocket back inner | -0.01 | 0.3 | -0.04 | 0.5 |

[a]Reference turns were detected with the motion capture system. SD, standard deviation.

## 3.3. Validation of turn speed median

In the supervised setting, the median (IQR) of turn speed median was 1.44 (1.2 – 1.7) rad/s when measured with the smartphone carried in the belt front location) and 1.48 (1.1 – 1.8) rad/s with the motion capture system. The results were similar across the six smartphone locations (Supplementary Table S3). In the unsupervised remote setting, turn speed median was 1.48 (1.2 – 1.7) rad/s. For additional details on turn speed median, turn duration and number of turns see Supplementary Tables S1 – S3.

### 3.3.1. Concordance and agreement between smartphone and reference system in the supervised setting

The turn speed median values derived from the smartphone showed high levels of concordance with the values derived from the motion capture system. This was consistent across disability groups (Figure 4) but showed some variation across smartphone locations: good concordance for the front (ICC[3,1] = 0.87) and back (ICC[3,1] = 0.88) belt locations, and excellent concordance for all pocket locations (ICC [3,1] range: 0.90 – 0.92).

High levels of agreement between the two measurement systems were also observed, as represented in the Bland-Altman plots (Supplementary Figure S1). Across all six smartphone wear locations, limits of agreement were between -0.41 rad/s and 0.43 rad/s, with no significant bias observed between the two systems (bias close to zero), except for the front pocket on the outer leg (bias [95% CI]: 0.11 [0.08; 0.13] (Table 4).

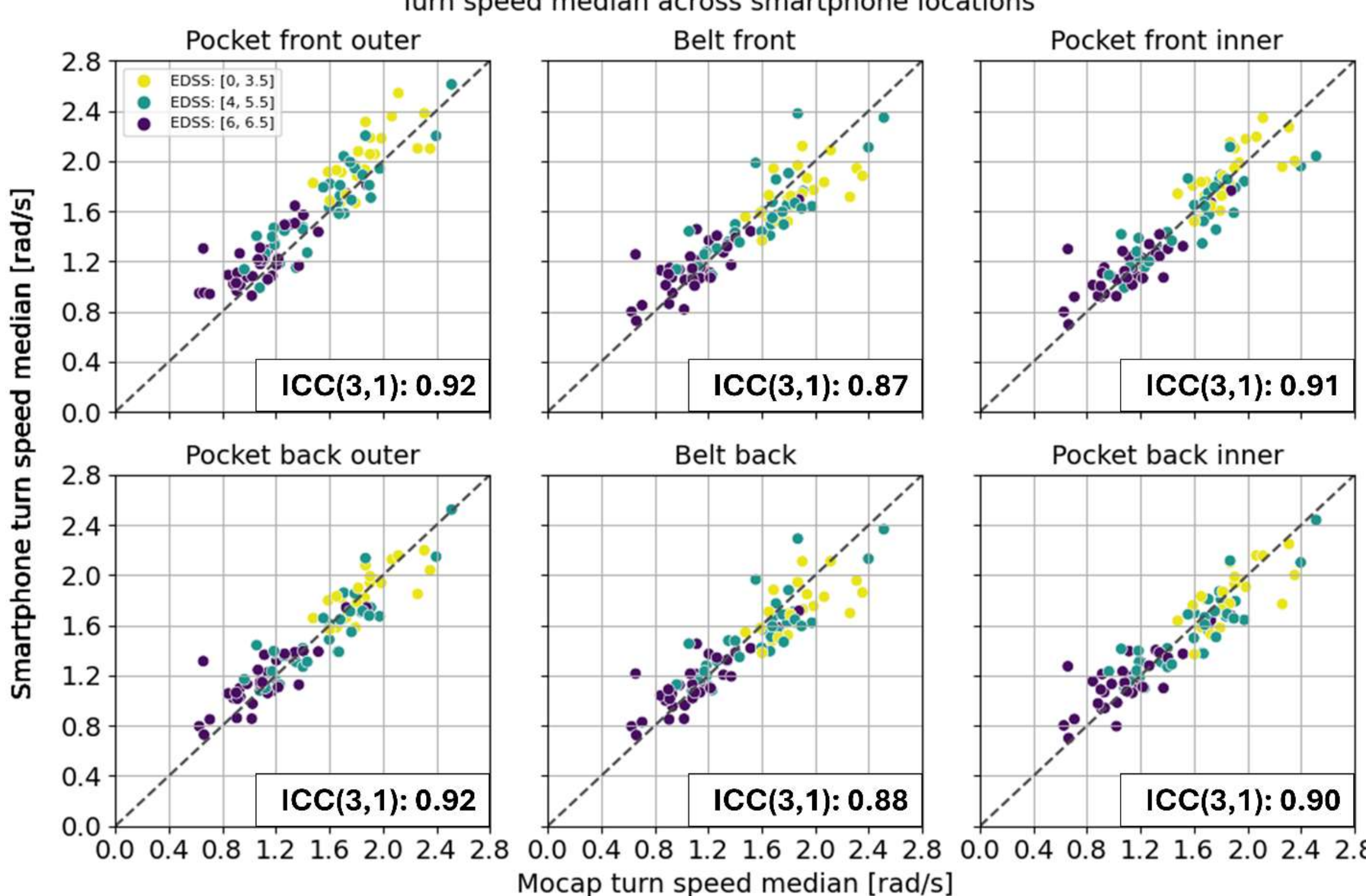

**Figure 4. Concordance plots for turn speed median measured with smartphone and motion capture systems during the supervised setting.**

ICC, intraclass correlation coefficient.

**Table 4. Agreement of turn speed median measured against motion capture system in the supervised setting.**

| | Turn speed, median (IQR) | Bias (95% CI) | Lower LoA (95% CI) | Upper LoA (95% CI) | ICC(3,1) (95% CI) |
|---|---|---|---|---|---|
| **Smartphone** | | | | | |
| Pocket front outer | 1.58 (0.7) | 0.11 (0.08; 0.13) | -0.22 (-0.26; -0.18) | 0.43 (0.39; 0.47) | 0.92 (0.89; 0.94) |
| Pocket back outer | 1.40 (0.6) | 0.01 (-0.02; 0.04) | -0.30 (-0.36; -0.25) | 0.32 (0.27; 0.38) | 0.92 (0.88; 0.95) |
| Belt front | 1.44 (0.5) | -0.03 (-0.06; 0.01) | -0.41 (-0.46; -0.35) | 0.36 (0.30; 0.41) | 0.87 (0.82; 0.91) |
| Belt back | 1.45 (0.6) | -0.04 (-0.07; -0.00) | -0.40 (-0.46; -0.35) | 0.33 (0.27; 0.39) | 0.88 (0.83; 0.92) |
| Pocket front inner | 1.42 (0.6) | 0.01 (-0.03; 0.04) | -0.33 (-0.39; -0.28) | 0.34 (0.29; 0.40) | 0.91 (0.87; 0.94) |
| Pocket back inner | 1.40 (0.6) | -0.01 (-0.03; 0.03) | -0.34 (-0.39; -0.30) | 0.33 (0.29; 0.38) | 0.90 (0.86; 0.94) |
| **Reference measurement system** | | | | | |
| Motion capture system | 1.48 (0.7) | N/A | N/A | N/A | N/A |

CI, confidence interval; ICC, intraclass correlation coefficient; IQR, interquartile range; LoA, limits of agreement; NA, not applicable.

### 3.3.2. Consistency across repeated assessments

In the unsupervised remote setting, patients performed an average (mean [95% CI]) of 11.6 (10.9 – 12.2) U-Turn tests over a 2-week period. For assessing repeatability, different aggregations were used which ranged from comparing 2 individual tests performed on different days, to comparing repeatability between two sets of 7 aggregated tests (Table 5). When comparing two individual tests, results showed good reproducibility (ICC[2,1]: 0.84 [95% CI: 0.81 – 0.88]). Additionally, SEM was 0.15 (95% CI: 0.13 – 0.16) rad/s, and MDC was 0.41 (95% CI: 0.37 – 0.46) rad/s, suggesting that a change of 23.5 deg/s (1 rad = 180 deg/π) can be detected when using non-aggregated measures of turn speed median. When aggregating as few as two tests, results showed excellent reproducibility (ICC > 0.90) with best results achieved for the aggregation of 7 tests (ICC = 0.97). With 7 aggregated tests, the SEM had its lowest value (0.06 [95% CI: 0.05 – 0.07] rad/s), and the MDC values were also the lowest (0.17 [95% CI: 0.14 – 0.20] rad/s), which is equivalent to an MDC of 9.7 deg/s (58.7% lower than for non-aggregated tests).

**Table 5. Test-rest reliability of turn speed median during the unsupervised, real-world setting.**

| Number of aggregated unsupervised UTT[a] | n[b] | ICC(2,1) (95% CI) | SEM (95% CI), rad/s | MDC (95% CI), rad/s |
|---|---|---|---|---|
| 1 | 91 | 0.84 (0.81; 0.88) | 0.15 (0.13; 0.16) | 0.41 (0.37; 0.46) |
| 2 | 89 | 0.91 (0.90; 0.93) | 0.11 (0.10; 0.11) | 0.30 (0.28; 0.32) |
| 3 | 85 | 0.92 (0.90; 0.93) | 0.10 (0.09; 0.11) | 0.29 (0.26; 0.31) |
| 4 | 80 | 0.95 (0.93; 0.96) | 0.09 (0.08; 0.09) | 0.24 (0.21; 0.26) |
| 5 | 72 | 0.94 (0.93; 0.95) | 0.09 (0.08; 0.09) | 0.24 (0.21; 0.26) |
| 6 | 62 | 0.95 (0.94; 0.96) | 0.08 (0.07; 0.08) | 0.21 (0.19; 0.23) |
| 7 | 26 | 0.97 (0.96; 0.98) | 0.06 (0.05; 0.07) | 0.17 (0.14; 0.20) |

[a]Measures of turn speed median obtained from individual unsupervised UTTs were aggregated by taking the median on a participant level. [b]The actual number of completed unsupervised UTT differed from patient to patient and, consequently, the sample size decreases when increasing the number of aggregated UTTs. ICC, intraclass correlation coefficient; MDC, minimal detectable change; PwMS, people with multiple sclerosis; SEM, standard error of mean.

### 3.3.3. Correlation with clinical outcomes

Turn speed median derived from a smartphone carried in a belt worn at the waist level (belt front), in an unsupervised remove setting, showed strong correlations with clinical measures of gait impairment such as the T25FW time ($\rho = -0.80$), and the Ambulation score ($\rho = -0.72$), and to a lesser extent with a measure of patient-perceived walking impairment (MSWS-12 transformed score, $\rho = -0.65$; Figure 5, panels A, B and C). Patients with unrestricted walking (Ambulation score 0 or 1, n = 41) had a significantly higher turn speed median ($p < 0.0001$) than patients with restricted walking but not requiring assistance (ambulation score 2-5, n = 49; Figure 6, panel A). Similarly, patients not requiring a walking aid (n = 39) had a significantly higher turn speed median (p <

0.0001) than those requiring an aid (n = 49; Figure 6, panel B). Correlation with overall disability as defined by EDSS was also strong (ρ = -0.74; Figure 5, panel D), with significant differences in turn speed across different EDSS groups (EDSS [0, 3.5] vs [4.0, 5.5], vs [6.0, 6.5], p < 0.001; Figure 6, panel C). Interestingly, correlation was lowest with the ABC scale, a self-report measure of balance confidence in performing various activities without losing balance or experiencing a sense of unsteadiness (ρ = -0.58; Figure 5 panel E). Finally, turn speed was significantly lower in patients who had reported to have had at least one fall in the last 3 months before enrolling into the study (p < 0.0001, Figure 6 panel D).

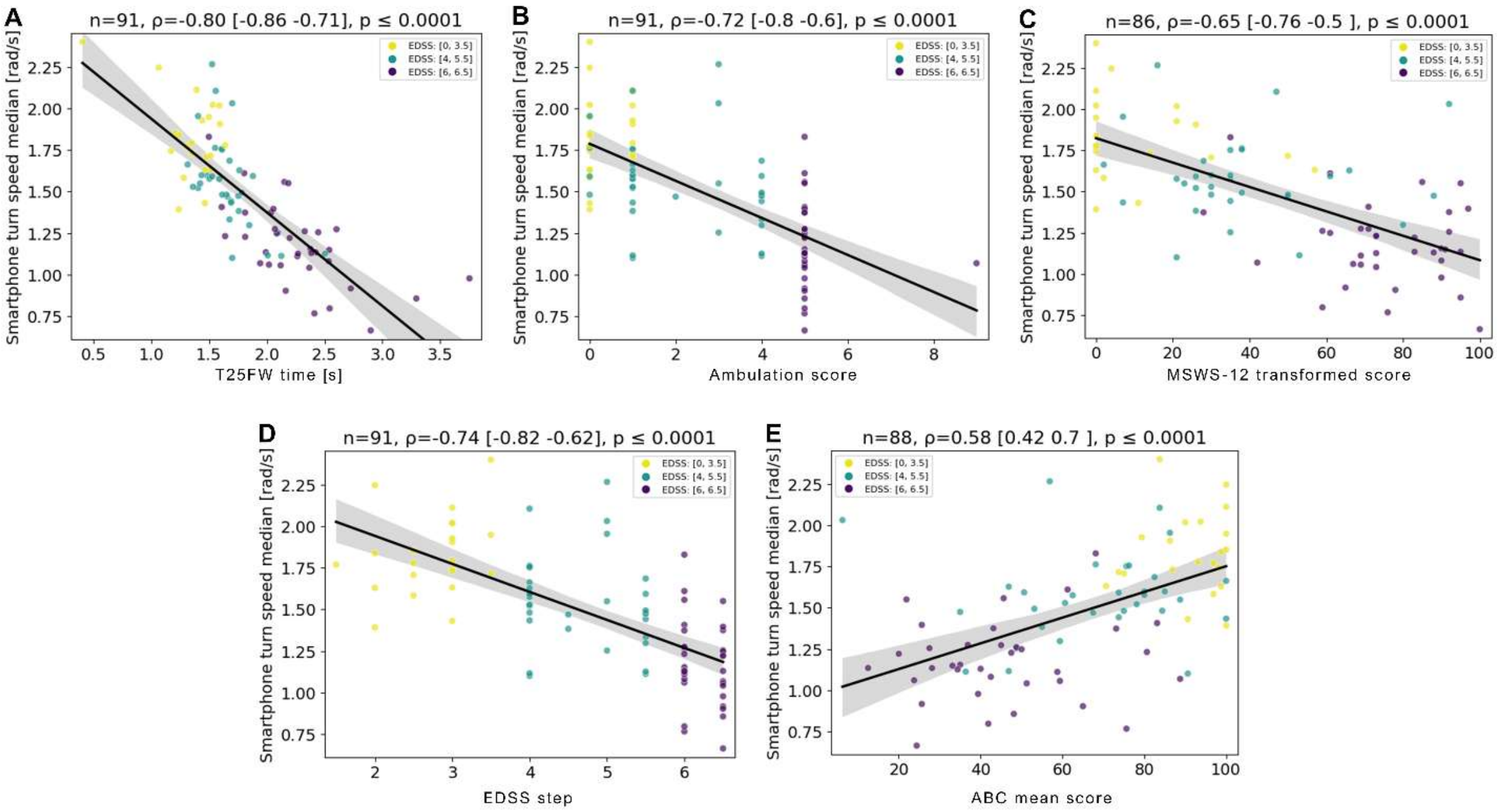

**Figure 5. Clinical correlations of turn speed median measured in the unsupervised, real-world setting.** Cross-sectional Spearman rank correlation between unsupervised UTT (turn speed median) and T25FW (A), Ambulation score (B), MSWS-12 transformed score (C), EDSS (D), and ABC mean score (E). Measures of turn speed median obtained from individual unsupervised UTTs were aggregated by taking the median on a participant level, resulting in one measure per participant. ABC, Activities-specific Balance Confidence scale; EDSS, Expanded Disability Status Scale; MSWS-12, 12-item Multiple Sclerosis Walking Scale; T25FW, Timed 25-Foot Walk.

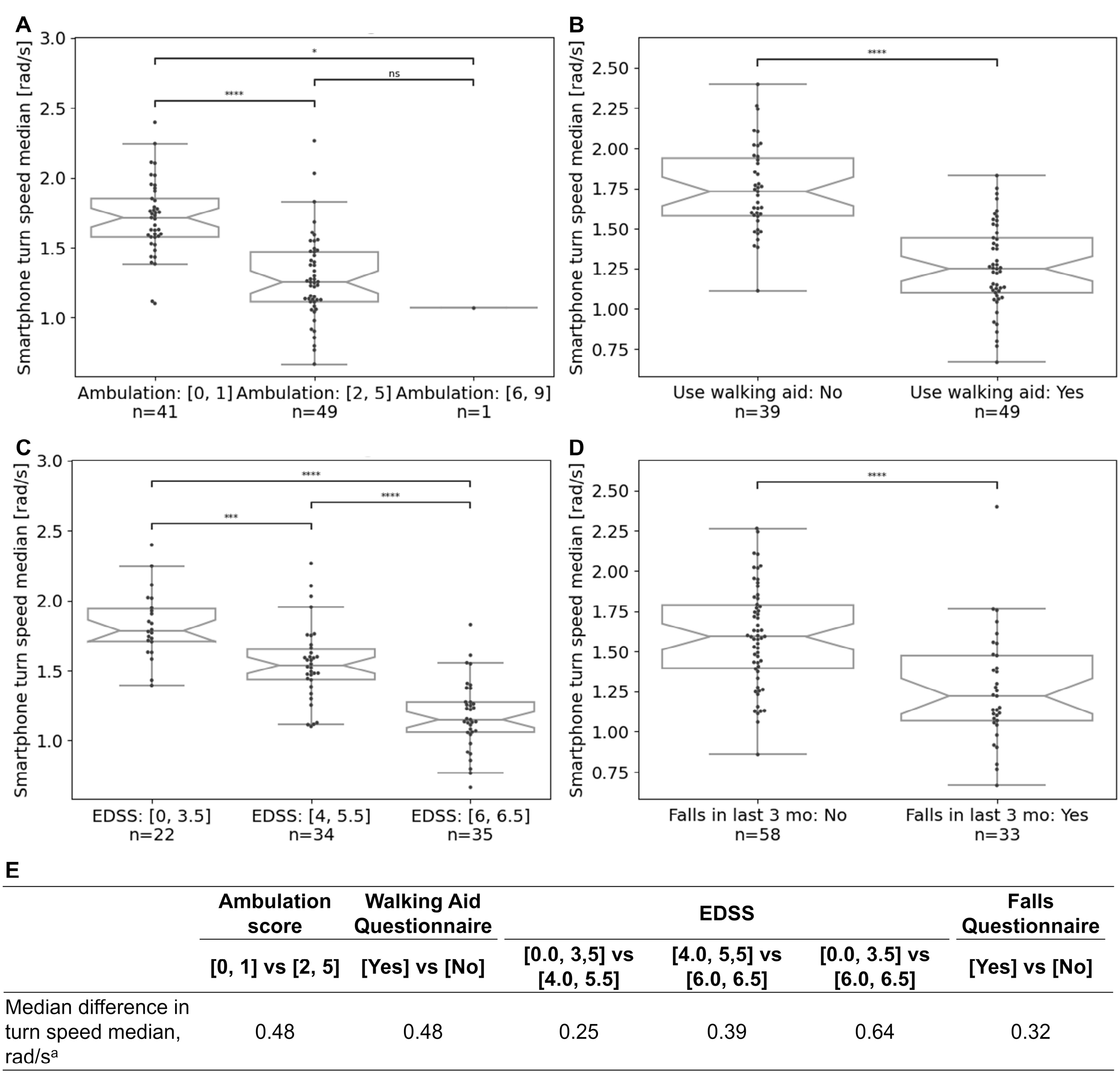

| | Ambulation score | Walking Aid Questionnaire | EDSS | | | Falls Questionnaire |
|---|---|---|---|---|---|---|
| | [0, 1] vs [2, 5] | [Yes] vs [No] | [0.0, 3,5] vs [4.0, 5.5] | [4.0, 5,5] vs [6.0, 6.5] | [0.0, 3.5] vs [6.0, 6.5] | [Yes] vs [No] |
| Median difference in turn speed median, rad/s[a] | 0.48 | 0.48 | 0.25 | 0.39 | 0.64 | 0.32 |

**Figure 6. Turn speed median across different patient groups, measured in the unsupervised, real-world setting.** Differences on the unsupervised UTT between groups of PwMS with different levels of disease burden measured by Ambulation score (A), Walking Aid Questionnaire (B) EDSS (C), and Falls Questionnaire (D). Mean differences between the patient groups are reported in panel (E). [a]Measures of turn speed median obtained from individual unsupervised UTTs were aggregated by taking the median on a participant level, resulting in one measure per participant. ****p<0.0001; ***p<0.001; **p<0.01; *p<0.05; ns, not significant.

EDSS, Expanded Disability Status Scale.

# 4. Discussion

Gait and balance impairments profoundly affect the safety, independence, and quality of life of PwMS (LaRocca 2011; Prosperini and Castelli 2018; Thompson et al. 2018a; Yildiz 2012). Traditional clinical assessments, while valuable, are performed infrequently in healthcare settings that may not reflect a person's real-world functioning (Sebastião et al. 2016; Cattaneo et al. 2006). This study bridges the gap by establishingthe analytical and cross-sectional clinical validity of the smartphone-based UTT as a tool for frequent and remote assessment of dynamic balance in MS. Our findings demonstrate high agreement with a laboratory-grade, gold-standard motion capture system, excellent reproducibility in unsupervised, real-world settings, and strong correlations with established clinical indicators of MS-related disability.

## 4.1. The smartphone-based UTT detects U-turns accurately

The foundation of any digital measure is its ability to accurately and consistently capture the targeted behavior. In this study we assessed, for the first time, the core performance of the smartphone-based turn detection algorithm against a well-established reference system. The algorithm is highly effective at both correctly identifying turns when they occur (sensitivity or recall > 95%) and avoiding the incorrect classification of non-turn movements as turns (precision > 96%), resulting in F1 scores that exceed 95%. While the sensitivity in our study exceeded the sensitivity (90%) reported by El-Gohary et al. in a population of people with Parkinson’s disease, there are some critical differences between the two algorithms, including the minimum angle threshold for detecting a turn (90 degrees in our study vs 45 degrees in El-Gohary et al.) (El-Gohary et al. 2013). Furthermore, the turn detection algorithm showed high temporal precision, with a near perfect overlap in the start and end times of each turn detected by reference systems. Mean temporal errors correspond to approximately 5% of the mean duration of turns.

## 4.2. The smartphone-based UTT show excellent concordance and agreement with reference systems

Smartphone-derived turn speed median showed good-to-excellent concordance and agreement with measurements obtained from a motion capture system (ICC[3,1]: 0.87 – 0.92). Notably, both the observed bias ( -0.04 to 0.01 rad/s) and limits of agreement (approximately ± 0.45 rad/s) are much smaller than the expected individual variability (within-subject IQR = 0.60 rad/s), suggesting that measurement error is well within the expected range of behavioral variation.

## 4.3. The smartphone-based UTT is robust to wear location

A key practical finding was the consistency of the results on turn speed median across various smartphone wear locations, including different belt and trouser pocket locations. For a test designed for remote, unsupervised use, this flexibility is paramount. It accommodates user preference and acknowledges that day-to-day placement may vary, removing a significant potential barrier to data quality and adherence. While our analysis did identify a small, statistically significant positive bias when the phone was in the front pocket of the outer leg during a turn, this was likely due to the combined rotational movement of the body and the swinging motion of the leg. Critically, the magnitude of this bias (0.11 rad/s) is negligible considering that it is smaller than the MDC (Table 5). All other smartphone wear locations showed no significant bias with the motion capture system (bias close to zero with confidence intervals that cross zero).

## 4.4. The smartphone-derived UTT shows excellent reproducibility at home without supervision

For a measure to be useful for monitoring patients with MS over time, it must be highly reproducible. Our results show that the smartphone-based UTT shows excellent consistency across repeated measurements when performed by PwMS in their own homes without supervision. An ICC(2,1) value greater than 0.90 was achieved by aggregating the results from as few as two separate tests. This is a significant finding, as it indicates that by combining just a couple of assessments, we can filter out day-to-day variability and obtain a stable, reliable estimate of a patient's functional ability.

The high reliability translates into a greater sensitivity to detect genuine changes in a patient's condition. As more tests are aggregated, the measurement error decreases, leading to a smaller MDC, the minimum change that can be considered real and not just random variation (Jaeschke et al. 1989; Napiórkowski et al. 2024). This improved reproducibility represents an important advancement over precursor versions of the UTT (ICCs 0.83 – 0.87) (Cheng et al. 2021; Montalban et al. 2022), likely resulting from refinements to the processing algorithms and the inclusion of a more diverse MS cohort in our study, which increased between-subject variability (a factor that naturally enhances the statistical reliability [ICC] of a clinical measure).

## 4.5. The smartphone-derived UTT is clinically meaningful

The UTT demonstrated clear clinical relevance by correlating strongly with established measures of MS severity, including T25FW gait speed ($\rho$ = - 0.80), EDSS ($\rho$ = - 0.74), and the Ambulation score ($\rho$ = - 0.72), whereas previous studies showed correlations with EDSS and T25FW of moderate strength (Cheng et al. 2021; Chitnis et al. 2019; Montalban et al. 2022). The stronger correlations observed in this study may be explained by the improved UTT algorithms used in this analysis and the broader MS population enrolled in this study. Notably, the UTT exhibited a less pronounced floor effect than the T25FW (Figure 5A), indicating potential superior sensitivity in mildly impaired patients.

These relationships confirm that the UTT does not measure an isolated biomechanical variable, but that it instead captures an aspect of dynamic balance that is intrinsically linked to a patient's overall disability level and walking ability as assessed in the clinic (Anastasi et al. 2023; Peebles et al. 2018). Moreover, measures of turn speed have been shown to be independent predictors of patient-reported balance confidence and gait impairment and can augment the assessment of ambulation in PwMS (Adusumilli et al. 2018).

The UTT could effectively differentiate among PwMS with varying levels of impairment. As expected, PwMS with more advanced disease (higher EDSS), those who required a walking aid, and those with more limited ambulation had significantly slower turn speeds. Additionally, and importantly, turn speed median allowed to discriminate between individuals who had fallen in the three months prior to the study and those who had not. Given that falls are a frequent and dangerous complication of MS (Coote et al. 2020; Gunn et al. 2013; Jawad et al. 2023), this finding suggests the UTT has considerable potential as a future tool for remotely monitoring fall risk, which could enable more timely and targeted interventions.

## 4.6. Limitations

Several limitations should be considered when interpreting these findings. First, it is important to acknowledge that the definition of a "turn" can differ slightly between measurement technologies (e.g., marker-based vs. inertial sensor-based), which can introduce small temporal discrepancies.

For example, onset of turns can be defined as the time point at which the rotation of the trunk exceeds a preset threshold (El-Gohary et al. 2013) or when the direction of gait measured through step landmarks changes (Rehman et al. 2020). Hence, the temporal errors of turn onset and turn end need to be interpreted with caution. Second, we were unable to assess the UTT's agreement with a reference system in the remote setting. The two-sensor wearable system used for this purpose was insufficient for accurately detecting turn timing, largely because it lacked a sensor near the body's center of mass to capture torso rotation (Shah et al. 2020). However, the strong agreement in the lab and the lack of significant bias between the supervised and unsupervised settings provide confidence that the test remains accurate when used at home. A further, but minor limitation is that only a single smartphone model was used in this study. While this may limit the generalizability of the findings, previous exploratory analyses demonstrated a high degree of measurement equivalence of turn speed across a wide range of Android and iOS smartphone models (Kriara et al. 2025). Finally, the current turn detection algorithm had difficulty processing the very slow, prolonged turns of a few participants with an EDSS of 6.5. In these cases, turns were often so slow and fragmented that the angular velocity fell below the algorithm's detection threshold. Consequently, while the UTT is highly effective for most PwMS, these findings suggest a functional ceiling in which the signal-to-noise ratio in gyroscope data becomes insufficient for automated segmentation. Future iterations of the algorithm could incorporate lower, more sensitive thresholds specifically for highly impaired cohorts.

## 4.7. Conclusions

In conclusion, this comprehensive validation study demonstrates that the UTT is an analytically valid and clinically relevant smartphone-based assessment for dynamic balance in patients with MS. The test provides accurate and highly reliable measures of turn speed in both controlled laboratory and unsupervised home environments. The flexibility regarding smartphone placement enhances its real-world usability. The strong associations with key clinical measures of disability, walking impairment, and fall history confirm the clinical relevance of the UTT. These findings support the potential of the UTT as an analytically and cross-sectionally clinically valid, reliable, and sensitive tool for monitoring dynamic balance more frequently and more conveniently in the daily lives of patients in the context of clinical trials. To fully establish its utility, future work will need to investigate its sensitivity to longitudinal change.

## Acknowledgments

The authors would like to thank all study participants and their families. We would also like to thank the following (former) colleagues at F. Hoffmann-La Roche Ltd for their contributions to and support of the study: Matthias Bobst, Alan K. Bourke, Adrian Derungs, Sandro Fritz, Francisco Gavidia, Thomas Hahn, Petra Hauser, Wiktoria Kasprzyk, Dominik Kedziora, Kostas Kritsas, Hugo Le Gall, Dominik Lenart, Michael Lindemann, Florian Lipsmeier, Arnaud Mousley, Kathrin Müsch, Madalina Ogica, Emanuele Passerini, Grégoire Pointeau, and Elena Spyridou. The authors would also like to thank colleagues at Plymouth for their contributions and support to the study: Tanya King, Joanne Lind, Carol Lunn, and Maureen Pedder. Writing and editorial assistance for this manuscript was provided by Sven Holm, PhD, contractor for F. Hoffmann-La Roche Ltd.

## Disclosures

This research was funded by F. Hoffmann-La Roche Ltd, Basel, Switzerland. The study was designed jointly by the sponsors and investigators. Data were collected by the investigators and held and analyzed by the sponsor. All authors had full access to the data and approved its completeness and analysis. The authors had final responsibility for the decision to submit for publication.

## Data availability

Request for the data underlying this publication requires a detailed, hypothesis-driven statistical analysis plan that is collaboratively developed by the requester and company subject matter experts. Such requests should be directed to dbm.datarequest@roche.com for consideration. Anonymized records for individual patients across >1 data source external to Roche cannot, and should not, be linked due to a potential increase in the risk of patient reidentification.

## CRediT authorship contribution statement

**MP** Data Curation, Formal analysis, Methods, Software, Visualization, Writing – Original Draft, Writing – Review & Editing

**RK** Data Curation, Formal analysis, Methods, Software, Visualization, Writing – Review & Editing

**DS** Conceptualization, Data Curation, Formal analysis, Methods, Software, Visualization, Writing – Review & Editing

**LA** Conceptualization, Data Curation, Formal analysis, Methods, Software, Visualization, Writing – Review & Editing

**NN** Data Curation, Formal analysis, Software, Visualization, Writing – Review & Editing

**GGC** Data Curation, Investigation, Writing – Review & Editing

**LB** Data Curation, Investigation, Writing – Review & Editing

**PSG** Data Curation, Investigation, Writing – Review & Editing

**RH** Data Curation, Investigation, Writing – Review & Editing

**JF** Data Curation, Investigation, Writing – Review & Editing, Conceptualization

**JH** Data Curation, Investigation, Writing – Review & Editing, Conceptualization

**MZ** Conceptualization, Formal analysis, Methods, Project administration, Supervision, Writing – Review & Editing

**JM** Conceptualization, Data Curation, Investigation, Supervision, Writing – Review & Editing

**LC** Conceptualization, Funding acquisition, Formal analysis, Methods, Supervision, Writing – Review & Editing

**MDR** Conceptualization, Formal analysis, Methods, Project administration, Supervision, Writing – Original Draft, Writing – Review & Editing

## Declaration of competing interests

The author(s) declared the following potential conflicts of interest concerning the research, authorship, and/or publication of this article:

**MP** is a contractor for F. Hoffmann-La Roche Ltd.

**RK** is a contractor for F. Hoffmann-La Roche Ltd.

**DS** is a consultant for F. Hoffmann-La Roche Ltd.

**LA** is an employee of and a shareholder in F. Hoffmann-La Roche Ltd.

**NN** is a contractor for Roche Polska Sp. z o.o.

**GGC** is an employee of the University of Plymouth, which received research funding from F. Hoffmann-La Roche Ltd.

**LB** is an employee of the University of Plymouth, which received research funding for this study from F. Hoffmann-La Roche Ltd.

**PSG** is an employee of the University of Plymouth, which received research funding for this study from F. Hoffmann-La Roche Ltd.

**RH** received indirect funding from F. Hoffmann-La Roche Ltd via a University of Plymouth Enterprise Ltd contract.

**JF** is an employee of the University of Plymouth, which received research funding for this study from F. Hoffmann-La Roche Ltd.

**JH** or affiliated institutions have received either consulting fees, honoraria, support to attend meetings, clinical service support or research support from Acorda, Bayer Schering Pharma, Biogen Idec., Brickell Biotech, F. Hoffmann-La Roche Ltd, Global Blood Therapeutics, Sanofi-Genzyme, Merck-Serono, Novartis, Oxford Health Policy Forum, Teva and Vantia

**MZ** is an employee of and a shareholder in F. Hoffmann-La Roche Ltd.

**JM** is an employee of the University of Plymouth, which received research funding for this study from F. Hoffmann-La Roche Ltd.

**LC** is an employee of and a shareholder in F. Hoffmann-La Roche Ltd.

**MDR** is a contractor for F. Hoffmann-La Roche Ltd.

*Supplementary appendix to:*

# Analytical and Cross-Sectional Clinical Validity of a Smartphone-Based U-Turn Test in Multiple Sclerosis

## Remote study reference validation

During the remote testing period, participants were provided with two body-worn IMUs manufactured by Gait Up (Lausanne, Switzerland, Physilog 6® model). Sensor data were collected at a sampling frequency of 128 Hz, with both IMU sensors attached to the participants' feet. This two-sensor setting was chosen over a three-sensor setting to reduce patient burden and aimed primarily to serve as a reference system for gait. However, we decided not to use it as a reference for turn speed due to poor performance on validation of its algorithm against the motion capture reference system and inadequate in timing of the turn begin and end which caused incorrect estimation of turn speed. When comparing a 3-IMU sensor setup to a 2-IMU setup, it's essential to note that the additional sensor close to the body's center of mass plays a critical role in capturing accurate rotational and postural data. Technical validation of 2-IMU sensors against the motion capture reference system resulted in a mean F1 score of 88.5 (SD: 26.1; 95% CI: 83.1 – 93.9). The mean error was -0.19 s (SD: 0.6 s) for the detection of turn onset and 0.19 s (SD: 0.6) s for the detection of turn end. The median turn speed provided by the Gait Up 2-IMU sensor set-up algorithm showed higher variability than the motion capture system (Supplementary Figure S2), with moderate ICC(3,1) of 0.63 (95% CI: 0.55 – 0.71).

# Supplementary tables and figures

**Supplementary Table S1. Descriptive statistics of number of turns[a]**

| Experimental setup | Smartphone wear location | n | Missing * | Number of turns | | | | | | | | | | |
|---|---|---|---|---|---|---|---|---|---|---|---|---|---|---|
| | | | | **Mean (95% CI)** | **SD** | **Min** | **Max** | **P05** | **Q1** | **Med** | **Q3** | **P95** | **IQR** | **ws-IQR** |
| Supervised UTT | | | | | | | | | | | | | | |
| Smartphone | Belt back | 96 | 0 | 11.4 (10.4; 12.4) | 4.9 | 0 | 22 | 3.5 | 8 | 12.0 | 15 | 19 | 7 | N/A[a] |
| | Belt front | 96 | 0 | 11.4 (10.5; 12.4) | 4.9 | 0 | 22 | 3.8 | 8 | 12.0 | 15 | 19 | 7 | N/A[a] |
| | Pocket back outer | 96 | 0 | 11.5 (10.5; 12.4) | 4.9 | 0 | 21 | 3.8 | 8 | 12.0 | 15 | 19 | 7 | N/A[a] |
| | Pocket front outer | 96 | 0 | 11.4 (10.4; 12.4) | 5.0 | 0 | 22 | 2.8 | 8 | 12.0 | 15 | 19 | 7 | N/A[a] |
| | Pocket back inner | 96 | 0 | 11.5 (10.5; 12.4) | 4.9 | 0 | 22 | 3.8 | 8 | 12.0 | 15 | 19 | 7 | N/A[a] |
| | Pocket front inner | 96 | 0 | 11.4 (10.4; 12.4) | 4.9 | 0 | 21 | 2.8 | 8 | 11.5 | 15 | 19 | 7 | N/A[a] |
| Motion capture reference system | N/A | 96 | 0 | 11.3 (10.3; 12.2) | 4.7 | 1 | 21 | 3.0 | 8 | 11.0 | 15 | 19 | 7 | N/A[a] |
| Unsupervised, remote UTT | | | | | | | | | | | | | | |
| Smartphone[b] | Belt | 91 | 5 | 8.48 (7.7; 9.2) | 3.68 | 3 | 19 | 3.5 | 5.5 | 8.0 | 11 | 15 | 6 | 5 |

[a]Each participant completed only one supervised UTT; hence, no ws-IQR could be computed. [b]Measures of turn speed median obtained from individual unsupervised UTTs were aggregated by taking the median on a participant level, resulting in one measure per participant. CI, confidence interval; IQR, interquartile range; Med, median; N/A, not applicable as each participant completed only one supervised UTT; P05, $5^{th}$ percentile; P95, $95^{th}$ percentile; SD, standard deviation; UTT, U-Turn Test; ws-IQR, within-subject interquartile range.

**Supplementary Table S2. Descriptive statistics of turn duration median.**

| Experimental setup | Smartphone wear location | n | Missing | Turn duration median, s | | | | | | | | | | |
|---|---|---|---|---|---|---|---|---|---|---|---|---|---|---|
| | | | | **Mean (95% CI)** | **SD** | **Min** | **Max** | **P05** | **Q1** | **Med** | **Q3** | **P95** | **IQR** | **ws-IQR** |
| Supervised UTT | | | | | | | | | | | | | | |
| Smartphone | Belt back | 93 | 3 | 2.2 (2.1; 2.3) | 0.5 | 1.2 | 3.9 | 1.4 | 1.9 | 2.1 | 2.6 | 2.9 | 0.7 | N/A[a] |
| | Belt front | 93 | 3 | 2.2 (2.1; 2.3) | 0.5 | 1.2 | 4.0 | 1.4 | 1.9 | 2.1 | 2.5 | 2.9 | 0.6 | N/A[a] |
| | Pocket back outer | 93 | 3 | 2.2 (2.2; 2.3) | 0.5 | 1.3 | 3.8 | 1.6 | 1.9 | 2.1 | 2.6 | 2.9 | 0.7 | N/A[a] |
| | Pocket front outer | 93 | 3 | 2.1 (2.0; 2.2) | 0.4 | 1.2 | 3.3 | 1.5 | 1.8 | 2.0 | 2.5 | 2.9 | 0.7 | N/A[a] |
| | Pocket back inside | 93 | 3 | 2.2 (2.1; 2.3) | 0.5 | 1.3 | 3.6 | 1.5 | 1.8 | 2.1 | 2.5 | 2.9 | 0.7 | N/A[a] |
| | Pocket front inside | 93 | 3 | 2.2 (2.1; 2.3) | 0.5 | 1.4 | 3.8 | 1.5 | 1.9 | 2.1 | 2.6 | 3.1 | 0.7 | N/A[a] |
| Motion capture reference system | N/A | 96 | 0 | 2.6 (2.5; 2.8) | 1.7 | 1.3 | 13.9 | 1.4 | 1.8 | 2.2 | 2.8 | 4.8 | 1.1 | N/A[a] |
| Unsupervised, remote UTT | | | | | | | | | | | | | | |
| Smartphone[b] | Belt | 91 | 5 | 2.1 (2.0; 2.2) | 0.4 | 1.3 | 3.4 | 1.5 | 1.8 | 2.0 | 2.3 | 2.8 | 0.5 | 0.6 |

[a]Each participant completed only one supervised UTT; hence, no ws-IQR could be computed. [b]Measures of turn speed median obtained from individual unsupervised UTTs were aggregated by taking the median on a participant level, resulting in one measure per participant. CI, confidence interval; IQR, interquartile range; Med, median; N/A, not applicable as each participant completed only one supervised UTT; P05, $5^{th}$ percentile; P95, $95^{th}$ percentile; SD, standard deviation; UTT, U-Turn Test; ws-IQR, within-subject interquartile range.

**Supplementary Table S3. Descriptive statistics of turn speed median**

| Experimental setup | Smartphone wear location | n | Missing, n | Turn speed median, rad/s | | | | | | | | | | |
|---|---|---|---|---|---|---|---|---|---|---|---|---|---|---|
| | | | | Mean (95% CI) | SD | Min | Max | P05 | Q1 | Med | Q3 | P95 | IQR | ws-IQR |
| Supervised UTT | | | | | | | | | | | | | | |
| Smartphone | Pocket front outer | 93 | 3 | 1.58 (1.49; 1.66) | 0.42 | 0.9 | 2.6 | 1.0 | 1.2 | 1.58 | 1.9 | 2.3 | 0.7 | N/A[a] |
| | Pocket back outer | 93 | 3 | 1.48 (1.41; 1.56) | 0.38 | 0.7 | 2.5 | 0.9 | 1;2 | 1.40 | 1.7 | 2.1 | 0.6 | N/A[a] |
| | Belt front | 93 | 3 | 1.45 (1.38; 1.52) | 0.35 | 0.7 | 2.4 | 0.9 | 1.2 | 1.44 | 1.7 | 2.0 | 0.5 | N/A[a] |
| | Belt back | 93 | 3 | 1.44 (1.37; 1.51) | 0.35 | 0.7 | 2.4 | 0.9 | 1.1 | 1.45 | 1.7 | 2.0 | 0.6 | N/A[a] |
| | Pocket front inner | 93 | 3 | 1.48 (1.40; 1.56) | 0.39 | 0.7 | 2.4 | 0.9 | 1.2 | 1.42 | 1.8 | 2.1 | 0.6 | N/A[a] |
| | Pocket back inner | 93 | 3 | 1.47 (1.39; 1.54) | 0.37 | 0.7 | 2.4 | 0.9 | 1.2 | 1.40 | 1.7 | 2.1 | 0.6 | N/A[a] |
| Motion capture reference system | N/A | 96 | 0 | 1.47 (1.39; 1.56) | 0.43 | 0.6 | 2.5 | 0.9 | 1.1 | 1.48 | 1.8 | 2.2 | 0.7 | N/A[a] |
| Unsupervised, remote UTT | | | | | | | | | | | | | | |
| Smartphone[b] | Belt | 91 | 5 | 1.47 (1.40; 1.55) | 0.36 | 0.7 | 2.4 | 0.9 | 1.2 | 1.48 | 1.7 | 2.1 | 0.5 | 0.6 |

[a]Each participant completed only one supervised UTT; hence, no ws-IQR could be computed. [b] Measures of turn speed median obtained from individual unsupervised UTTs were aggregated by taking the median on a participant level, resulting in one measure per participant. CI, confidence interval; IQR, interquartile range; Med, median; N/A, not applicable as each participant completed only one supervised UTT; P05, $5^{th}$ percentile; P95, $95^{th}$ percentile; SD, standard deviation; UTT, U-Turn Test; ws-IQR, within-subject interquartile range.

**Supplementary Table S4. Descriptive statistics on precision for turn detection algorithm applied during the supervised setting.**

| Smartphone wear location | Distribution of precision scores across PwMS[a] | | | | | | | | |
|---|---|---|---|---|---|---|---|---|---|
| | **Mean (95% CI)** | **SD** | **Min** | **Max** | **P05** | **Q1** | **Median** | **Q3** | **P95** |
| Pocket front outer | 96.4 (92.91; 99.96) | 17.6 | 0.0 | 100 | 91.5 | 100 | 100 | 100 | 100 |
| Pocket back outer | 96.2 (92.62; 99.73) | 17.7 | 0.0 | 100 | 81.7 | 100 | 100 | 100 | 100 |
| Belt front | 96.1 (92.54; 99.64) | 17.8 | 0.0 | 100 | 81.7 | 100 | 100 | 100 | 100 |
| Belt back | 96.1 (92.54; 99.64) | 17.8 | 0.0 | 100 | 81.7 | 100 | 100 | 100 | 100 |
| Pocket front inner | 96.2 (92.62; 99.73) | 17.7 | 0.0 | 100 | 81.7 | 100 | 100 | 100 | 100 |
| Pocket back inner | 96.2 (92.62; 99.73) | 17.7 | 0.0 | 100 | 81.7 | 100 | 100 | 100 | 100 |

[a]Precision scores are reported as a percentage. CI, confidence interval; SD, standard deviation; min, minimum value; max, maximum value; P05, 5th percentile; Q1, 1st quartile; Q3, 3rd quartile; P95, 95th percentile, PwMS, people with multiple sclerosis.

**Supplementary Table S5. Descriptive statistics on recall for turn detection algorithm applied during the supervised setting.**

| Smartphone wear location | Distribution of recall scores across PwMS[a] | | | | | | | | |
|---|---|---|---|---|---|---|---|---|---|
| | **Mean (95% CI)** | **SD** | **Min** | **Max** | **P05** | **Q1** | **Median** | **Q3** | **P95** |
| Pocket front outer | 95.2 (91.37; 99.03) | 19.2 | 0.0 | 100 | 66.7 | 100 | 100 | 100 | 100 |
| Pocket back outer | 95.7 (91.99; 99.35) | 18.4 | 0.0 | 100 | 83.3 | 100 | 100 | 100 | 100 |
| Belt front | 95.7 (91.98; 99.34) | 18.4 | 0.0 | 100 | 82.3 | 100 | 100 | 100 | 100 |
| Belt back | 96.1 (92.49; 99.75) | 18.2 | 0.0 | 100 | 89.4 | 100 | 100 | 100 | 100 |
| Pocket front inner | 95.2 (91.3; 99.09) | 19.5 | 0.0 | 100 | 62.5 | 100 | 100 | 100 | 100 |
| Pocket back inner | 95.6 (91.9; 99.28) | 18.4 | 0.0 | 100 | 79.2 | 100 | 100 | 100 | 100 |

[a]Recall scores are reported as a percentage. CI, confidence interval; SD, standard deviation; min, minimum value; max, maximum value; P05, 5th percentile; Q1, 1st quartile; Q3, 3rd quartile; P95, 95th percentile, PwMS, people with multiple sclerosis.

**Supplementary Table S6. F1 score statistics for turn detection algorithm applied during the supervised setting by EDSS group.**

| **Cohort and smartphone wear location** | **Distribution of F1 scores across PwMS[a]** | | | | | | | | |
|---|---|---|---|---|---|---|---|---|---|
| | **Mean (95% CI)** | **SD** | **Min** | **Max** | **P05** | **Q1** | **Median** | **Q3** | **P95** |
| Mild PwMS[b] | | | | | | | | | |
| Pocket front outer | 99.9 (99.7; 100) | 0.6 | 97.3 | 100 | 100 | 100 | 100 | 100 | 100 |
| Pocket back outer | 99.9 (99.7; 100) | 0.6 | 97.3 | 100 | 100 | 100 | 100 | 100 | 100 |
| Belt front | 99.9 (99.7; 100) | 0.6 | 97.3 | 100 | 100 | 100 | 100 | 100 | 100 |
| Belt back | 99.9 (99.7; 100) | 0.6 | 97.3 | 100 | 100 | 100 | 100 | 100 | 100 |
| Pocket front inner | 99.9 (99.7; 100) | 0.6 | 97.3 | 100 | 100 | 100 | 100 | 100 | 100 |
| Pocket back inner | 99.9 (99.7; 100) | 0.6 | 97.3 | 100 | 100 | 100 | 100 | 100 | 100 |
| Moderate PwMS[c] | | | | | | | | | |
| Pocket front outer | 99.3 (98.4; 100) | 2.6 | 89.7 | 100 | 92.9 | 100 | 100 | 100 | 100 |
| Pocket back outer | 99.3 (98.4; 100) | 2.6 | 89.7 | 100 | 92.9 | 100 | 100 | 100 | 100 |
| Belt front | 99.1 (98.3; 100) | 2.6 | 89.7 | 100 | 92.9 | 100 | 100 | 100 | 100 |
| Belt back | 99.3 (98.5; 100) | 2.5 | 89.7 | 100 | 94 | 100 | 100 | 100 | 100 |
| Pocket front inner | 99.4 (98.6; 100) | 2.4 | 89.7 | 100 | 97 | 100 | 100 | 100 | 100 |
| Pocket back inner | 99.2 (98.4; 100) | 2.6 | 89.7 | 100 | 92.3 | 100 | 100 | 100 | 100 |
| Severe PwMS[d] | | | | | | | | | |
| Pocket front outer | 89.5 (80.6; 98.4) | 28.1 | 0 | 100 | 0 | 100 | 100 | 100 | 100 |
| Pocket back outer | 90 (81.3; 98.8) | 27.5 | 0 | 100 | 0 | 100 | 100 | 100 | 100 |
| Belt front | 90 (81.3; 98.8) | 27.5 | 0 | 100 | 0 | 100 | 100 | 100 | 100 |
| Belt back | 90.5 (81.8; 99.3) | 27.5 | 0 | 100 | 0 | 100 | 100 | 100 | 100 |
| Pocket front inner | 88.9 (79.9; 97.8) | 28.3 | 0 | 100 | 0 | 100 | 100 | 100 | 100 |
| Pocket back inner | 89.9 (81.2; 98.7) | 27.5 | 0 | 100 | 0 | 100 | 100 | 100 | 100 |

[a]F1 scores are reported as a percentage. [b]PwMS with EDSS 0 – 3.5. [c]PwMS with EDSS 4.0 – 5.5. [d]PwMS with EDSS 6.0 – 6.5. CI, confidence interval; EDSS, Expanded Disability Status Scale; SD, standard deviation; min, minimum value; max, maximum value; P05, 5th percentile; Q1, 1st quartile; Q3, 3rd quartile; P95, 95th percentile, PwMS, people with multiple sclerosis.

**Supplementary Table S7. Precision score statistics for turn detection algorithm applied during the supervised setting by EDSS group.**

| Cohort and smartphone wear location | Distribution of precision scores across PwMS[a] | | | | | | | | |
|---|---|---|---|---|---|---|---|---|---|
| | **Mean (95% CI)** | **SD** | **Min** | **Max** | **P05** | **Q1** | **Median** | **Q3** | **P95** |
| Mild PwMS[b] | | | | | | | | | |
| Pocket front outer | 99.8 (99.3; 100) | 1.1 | 94.7 | 100 | 100 | 100 | 100 | 100 | 100 |
| Pocket back outer | 99.8 (99.3; 100) | 1.1 | 94.7 | 100 | 100 | 100 | 100 | 100 | 100 |
| Belt front | 99.8 (99.3; 100) | 1.1 | 94.7 | 100 | 100 | 100 | 100 | 100 | 100 |
| Belt back | 99.8 (99.3; 100) | 1.1 | 94.7 | 100 | 100 | 100 | 100 | 100 | 100 |
| Pocket front inner | 99.8 (99.3; 100) | 1.1 | 94.7 | 100 | 100 | 100 | 100 | 100 | 100 |
| Pocket back inner | 99.8 (99.3; 100) | 1.1 | 94.7 | 100 | 100 | 100 | 100 | 100 | 100 |
| Moderate PwMS[c] | | | | | | | | | |
| Pocket front outer | 98.9 (97.5; 100) | 4.3 | 81.2 | 100 | 94.5 | 100 | 100 | 100 | 100 |
| Pocket back outer | 98.9 (97.5; 100) | 4.3 | 81.2 | 100 | 94.5 | 100 | 100 | 100 | 100 |
| Belt front | 98.7 (97.2; 100) | 4.5 | 81.2 | 100 | 88.7 | 100 | 100 | 100 | 100 |
| Belt back | 98.7 (97.2; 100) | 4.5 | 81.2 | 100 | 88.7 | 100 | 100 | 100 | 100 |
| Pocket front inner | 98.9 (97.5; 100) | 4.3 | 81.2 | 100 | 94.5 | 100 | 100 | 100 | 100 |
| Pocket back inner | 98.9 (97.5; 100) | 4.3 | 81.2 | 100 | 94.5 | 100 | 100 | 100 | 100 |
| Severe PwMS[d] | | | | | | | | | |
| Pocket front outer | 92.1 (83.4; 100) | 27.3 | 0 | 100 | 0 | 100 | 100 | 100 | 100 |
| Pocket back outer | 91.4 (82.7; 100) | 27.4 | 0 | 100 | 0 | 100 | 100 | 100 | 100 |
| Belt front | 91.4 (82.7; 100) | 27.4 | 0 | 100 | 0 | 100 | 100 | 100 | 100 |
| Belt back | 91.4 (82.7; 100) | 27.4 | 0 | 100 | 0 | 100 | 100 | 100 | 100 |
| Pocket front inner | 91.4 (82.7; 100) | 27.4 | 0 | 100 | 0 | 100 | 100 | 100 | 100 |
| Pocket back inner | 91.4 (82.7; 100) | 27.4 | 0 | 100 | 0 | 100 | 100 | 100 | 100 |

[a]Precision scores are reported as a percentage. [b]PwMS with EDSS 0 – 3.5. [c]PwMS with EDSS 4.0 – 5.5. [d]PwMS with EDSS 6.0 – 6.5. CI, confidence interval; EDSS, Expanded Disability Status Scale; SD, standard deviation; min, minimum value; max, maximum value; P05, 5th percentile; Q1, 1st quartile; Q3, 3rd quartile; P95, 95th percentile, PwMS, people with multiple sclerosis.

**Supplementary Table S8. Recall score statistics for turn detection algorithm applied during the supervised setting by EDSS group.**

| Cohort and smartphone wear location | Distribution of recall scores across PwMS[a] | | | | | | | | |
|---|---|---|---|---|---|---|---|---|---|
| | Mean (95% CI) | SD | Min | Max | P05 | Q1 | Median | Q3 | P95 |
| Mild PwMS[b] | | | | | | | | | |
| Pocket front outer | 100 (100; 100) | 0 | 100 | 100 | 100 | 100 | 100 | 100 | 100 |
| Pocket back outer | 100 (100; 100) | 0 | 100 | 100 | 100 | 100 | 100 | 100 | 100 |
| Belt front | 100 (100; 100) | 0 | 100 | 100 | 100 | 100 | 100 | 100 | 100 |
| Belt back | 100 (100; 100) | 0 | 100 | 100 | 100 | 100 | 100 | 100 | 100 |
| Pocket front inner | 100 (100; 100) | 0 | 100 | 100 | 100 | 100 | 100 | 100 | 100 |
| Pocket back inner | 100 (100; 100) | 0 | 100 | 100 | 100 | 100 | 100 | 100 | 100 |
| Moderate PwMS[c] | | | | | | | | | |
| Pocket front outer | 99.7 (99.1; 100) | 1.9 | 88.9 | 100 | 100 | 100 | 100 | 100 | 100 |
| Pocket back outer | 99.7 (99.1; 100) | 1.9 | 88.9 | 100 | 100 | 100 | 100 | 100 | 100 |
| Belt front | 99.7 (99.1; 100) | 1.9 | 88.9 | 100 | 100 | 100 | 100 | 100 | 100 |
| Belt back | 100 (100; 100) | 0 | 100 | 100 | 100 | 100 | 100 | 100 | 100 |
| Pocket front inner | 100 (100; 100) | 0 | 100 | 100 | 100 | 100 | 100 | 100 | 100 |
| Pocket back inner | 99.6 (98.9; 100) | 2.1 | 87.5 | 100 | 100 | 100 | 100 | 100 | 100 |
| Severe PwMS[d] | | | | | | | | | |
| Pocket front outer | 88.2 (78.9; 97.5) | 29.2 | 0 | 100 | 0 | 100 | 100 | 100 | 100 |
| Pocket back outer | 89.4 (80.4; 98.3) | 28.3 | 0 | 100 | 0 | 100 | 100 | 100 | 100 |
| Belt front | 89.3 (80.3; 98.3) | 28.3 | 0 | 100 | 0 | 100 | 100 | 100 | 100 |
| Belt back | 90.2 (81.3; 99.1) | 28 | 0 | 100 | 0 | 100 | 100 | 100 | 100 |
| Pocket front inner | 87.9 (78.4; 97.3) | 29.7 | 0 | 100 | 0 | 100 | 100 | 100 | 100 |
| Pocket back inner | 89.2 (80.2; 98.2) | 28.3 | 0 | 100 | 0 | 100 | 100 | 100 | 100 |

[a]Recall scores are reported as a percentage. [b]PwMS with EDSS 0 – 3.5. [c]PwMS with EDSS 4.0 – 5.5. [d]PwMS with EDSS 6.0 – 6.5. CI, confidence interval; EDSS, Expanded Disability Status Scale; SD, standard deviation; min, minimum value; max, maximum value; P05, 5th percentile; Q1, 1st quartile; Q3, 3rd quartile; P95, 95th percentile, PwMS, people with multiple sclerosis.

**Supplementary Figure S1. Bland-Altman plots of turn speed median for different smartphone locations and turn speed measured with a reference system (motion capture system) obtained from the supervised UTT.[a]**

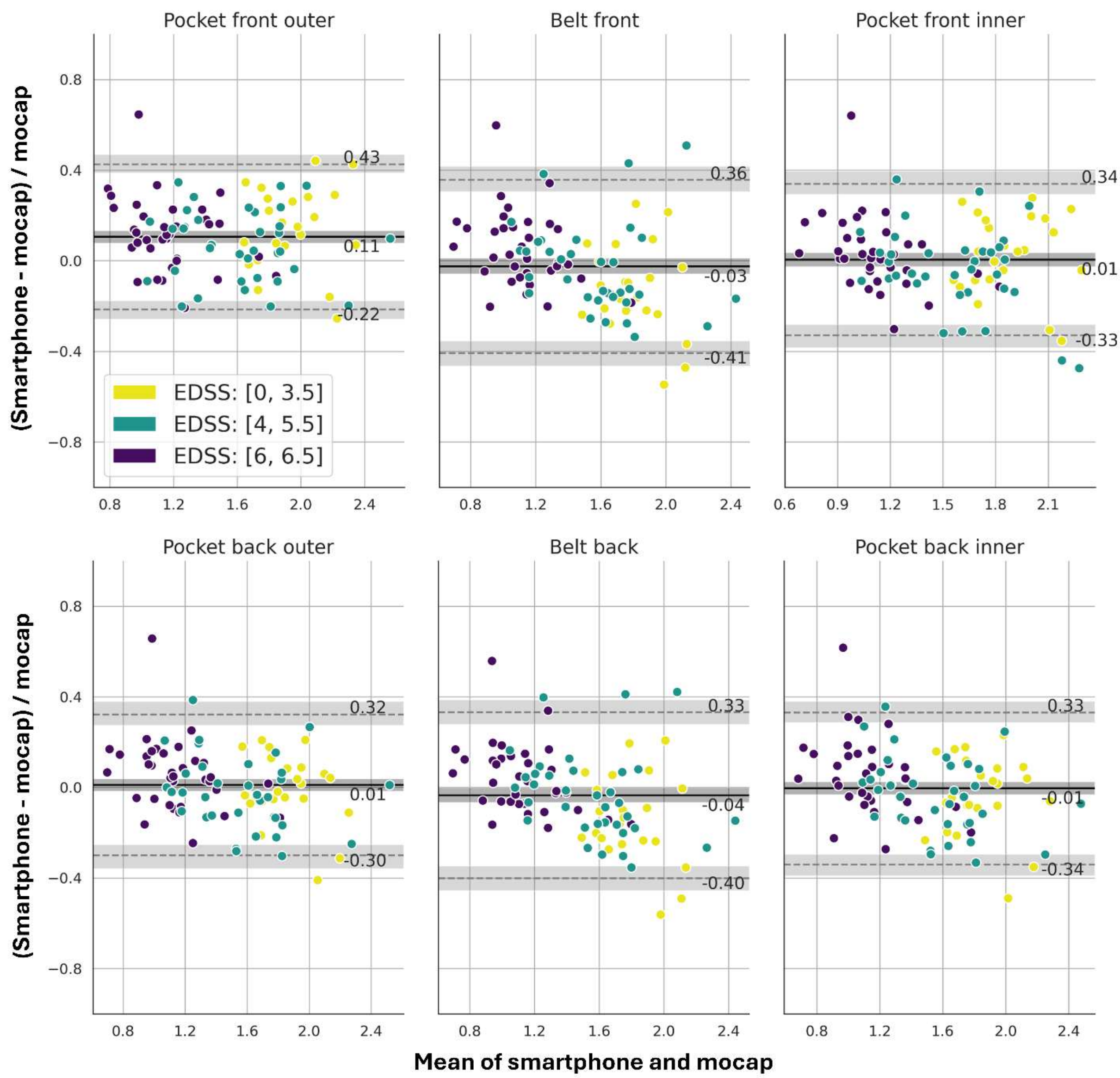

[a]All statistics presented on plots were obtained through a bootstrapping approach using 500 repetitions.

**Supplementary Figure S2. Scatter plot of turn speed measured during the supervised UTT with two IMU sensors (Gait Up) and a gold standard motion capture system.**

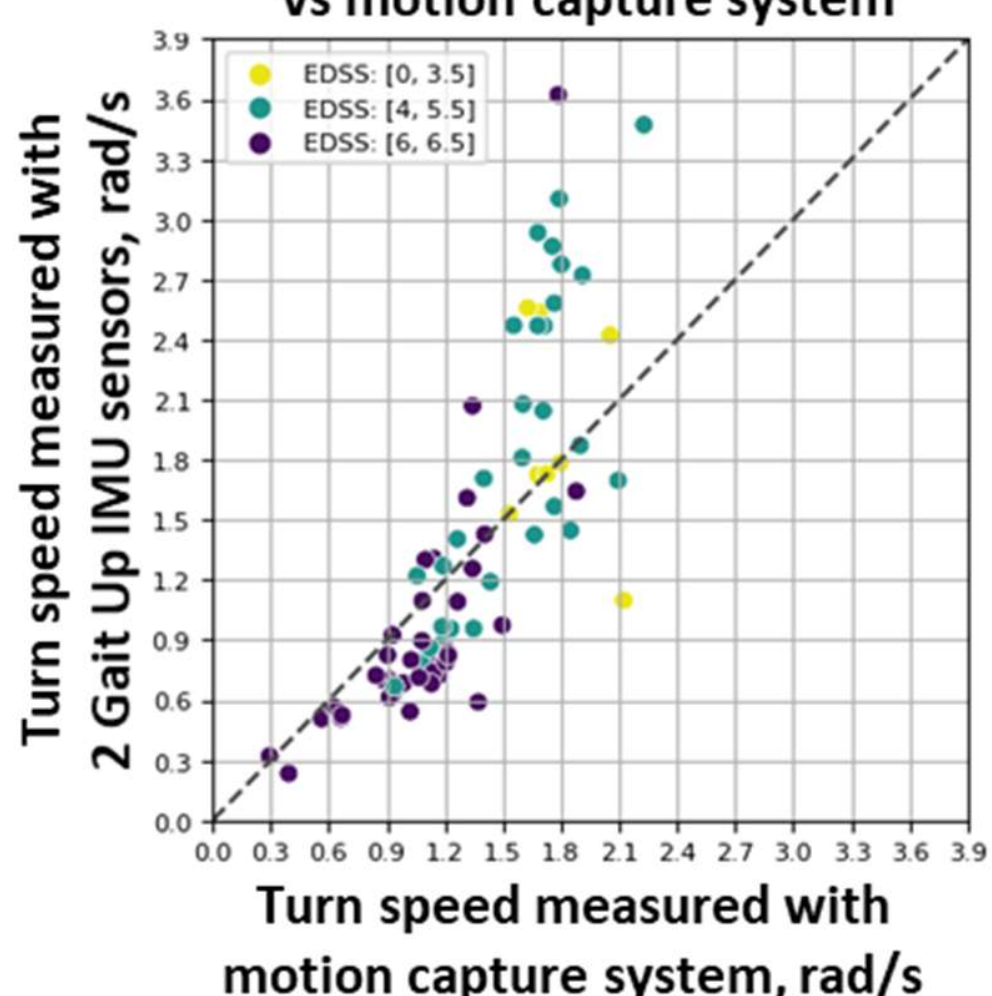